# Overcoming the challenges in controlled thermal deposition of organic diradicals

Tobias Junghoefer,[1] Nolan M. Gallagher,[2,†] Kubandiran Kolanji,[3,†] Erika Giangrisostomi,[4] Ruslan Ovsyannikov,[4] Thomas Chassé,[1] Martin Baumgarten,[3] Andrzej Rajca,[2] Arrigo Calzolari,[5] Maria Benedetta Casu[1*]

[1]Institute of Physical and Theoretical Chemistry, University of Tübingen, 72076 Tübingen, Germany

[2]Department of Chemistry, University of Nebraska, Lincoln, United States

[3]Max Planck Institute for Polymer Research, 55128 Mainz, Germany

[4]Helmholtz-Zentrum Berlin, 12489 Berlin, Germany

[5]CNR-NANO Istituto Nanoscienze, Centro S3, 41125 Modena, Italy

*Correspondence to: benedetta.casu@uni-tuebingen.de

Abstract.

We have demonstrated that it is possible to evaporate diradicals in a controlled environment obtaining thin films in which the diradical character is preserved. However, evaporation represents a challenge. The presence of two radical sites makes the molecules more reactive, even in case of very stable single radicals. We have explored the parameters that play a role in this phenomenon. Bulk formation thermodynamics and delocalisation of the unpaired electrons play the major role. The higher the formation energies of the crystal, the more difficult is the evaporation of intact radicals. The larger the delocalization, the more stable is the film exposed to air. The evaporation of different radicals, also with a larger number of radical sites can be successfully addressed considering our findings.

Technology based on quantum phenomena, such as entanglement and superposition, is taking the lead in various fields. Its fast development involves a strong multidisciplinary approach including a new technological vision in terms of applications and devices, new algorithms, and materials. The latter implies not only the strategic use of materials, but also the discovery and the design of new materials.[1] Among the materials candidate to play a role in emerging quantum technologies, organic materials and, particularly, radicals have recently attracted much attention.[2-5] Inert organic radicals and their derivatives are metal-free carbon-based molecules with one unpaired electron that are demonstrated to be stable enough to match technical requirements, such as evaporation and film processing.[6-10] However, the restriction to a single spin per molecule constitutes a limitation for many spintronic applications, so the synthesis and the control of multi-spin radicals would constitute a paramount step forward in the field. Diradicals have two unpaired electrons each localized in a specific chemical group in the molecules. The singlet and the triplet state in these molecules are in competition as a ground state. High spin diradicals with large energy gap between the triplet and the singles state, are interesting because of their potential use in applications, such as sensors, memories, and quantum gates.[11-15]

The implementation of the use of diradicals in real device has been hampered by the fact that their controlled evaporation and deposition onto a substrate was considered practically impossible to achieve because of their high reactivity.[16] Here we report the successful controlled evaporation of a diradical obtained fusing the nitronyl nitroxide (NN) radical to a derivative of the Blatter radical ($C_{26}H_{25}N_5O_2^{\bullet}$, NN-Blatter, Figure 1).[6] The NN and the Blatter single radicals are known to be stable against evaporation and they have good film forming properties, when fused to suitable substituents such as pyrene.[8, 10, 17] To tackle the problem of diradical evaporation, we have also evaporated a second diradical system, a derivative of the benzodithiophene decorated with two NN radicals ($C_{26}H_{32}N_4O_6S_2$, BTD-NN, Figure 5)[18] to identify the specific parameters that play a role for evaporation and film stability (Table 1) We have investigated the thin films by using X-ray based techniques and ab-initio calculations, focusing on diradical film processes and the consequent challenges that must be addressed to successfully evaporate and grow diradical thin films. The thin films were prepared in situ under ultrahigh vacuum (UHV) conditions by organic molecular beam deposition (OMBD), which allows tuning the preparation parameters to suit the film forming properties of the molecule.[19]

**Results and discussion:**

We used X-ray photoelectron spectroscopy (XPS) to investigate the thin films because it is proved an efficient tool for the investigation of organic radical thin films.[7] The method is element-sensitive, and allows insight into the stoichiometry of the film, and is sensitive towards the different chemical environments of atoms of the same element.[20]

We focus on the C 1s and N 1s core level spectra (for NN-Blatter in Figure 1) because the O 1s core level spectrum is the sum of the substrate ($SiO_2$/Si(111) wafers) and molecule signals, making the analysis unreliable. What is important is that the unpaired electrons are in the moieties containing nitrogen atoms. Therefore, the information on the radical is substantially delivered by the N 1s core level spectra. The NN-Blatter N 1s spectrum is composed of two main features. Two contributions originated by the photoelectrons emitted by the nitrogen atoms in the 2- and 4-position of the 1,2,4-benzotriazinyle moiety are visible centred around 400 eV. At higher binding energies, the emitted photoelectrons from the two chemically equivalent nitrogen atoms of the nitronyl nitroxide moiety combine with those of the nitrogen atom in the 1-position of the 1,2,4-benzotriazinyle moiety to form a broad peak with its highest intensity above 402 eV. The C 1s core level spectrum shows a main line at around 285 eV and a pronounced shoulder at higher binding energies. The main line contains the contributions from the emission of photoelectrons in the aromatic rings, as well as the contributions of carbon atoms bound to other carbon atoms and hydrogen atom (C-C, C-H, C-H$_3$). The shoulder is due to the signal of the photoelectrons emitted by carbon atoms bound to nitrogen atoms, which lie at higher binding energies due to the shift of charge towards the more electronegative nitrogen and the resulting decrease in electron density on the carbon atoms.

Figure 1 shows the thickness-dependent C 1s and N 1s core level spectra and it includes the best fit curves for the two core level spectra. The fit is forced by tight physical and chemical constraints (for details see the Supporting Information and Ref. [6, 8, 21]). The result indicates that the stoichiometry of the diradical film after evaporation and deposition corresponds to the theoretical values and, thus, that the diradical character of the molecule is maintained throughout the complete process.[6]

We also observe that there is a difference in the binding energy of the emitted photoelectrons when adopting a stepwise evaporation instead of a direct evaporation. The stepwise evaporation leads to a shift of the N 1s and C 1s spectra towards lower binding energies (Figure 2). This shift may be understood in terms of a change in the film morphology.[22] In fact, the increase of the thickness in organic thin films (e.g., monolayer vs. islands) leads to a less efficient screening of the core-hole, created upon photoemission, by the corresponding substrate mirror charge.[22-23] We explain our result as an "inverse" Stranski-Krastanov growth. In the Stranski-Krastanov growth mode, first, layer growth occurs, after islands nucleate on top of the layers. The NN-Blatter films grow following the Volmer−Weber (VW) growth mode, i.e., purely island growth in case of direct evaporation, as it is also confirmed by the atomic force microscopy images of the films that clearly show island formation (Figure 2).[6] Once the island are formed during deposition, a successive evaporation adds molecules on the substrate left free from the first deposition. This leads to contributions in the XPS spectra due to photoelectrons coming from thinner regions of the assembly, that is, closer to the substrate and characterised by a more efficient screening of the core-hole (Figure 2). Note that the screening of the core-hole by an image charge is observable at the organic/metal-oxides interface, however, its effect is weaker than in metal substrates.[24-25] We also observe that the shift is non-rigid (0.4 eV for the N 1s and 0.2 eV for the C 1s main line, respectively). Non-rigid shifts may have different reasons: the image charge of the core-hole originates in elements at different height from the substrate because the molecules are not completely planar on the substrate or in case of a fractional charge transfer from the substrate to the adsorbate in physisorption.[26] To investigate the strength of the interaction between NN-Blatter and the substrate, we performed an annealing experiment and we measured near-edge x-ray fine structure (NEXAFS) spectroscopy. In general, the island growth mode in thin films is driven by the fact that the intramolecular interaction is stronger than the interaction between the adsorbate and the substrate that cannot hence act as a template.[27] The polycrystalline NN-Blatter forms a one-dimensional spin-1 ($S = 1$) chain of radicals with intrachain antiferromagnetic (AFM) coupling of $J/k = -14$ K.[6] This is the strongest intrachain antiferromagnetic coupling among all studied one-dimensional spin-1 chains of organic radicals.[6] The intramolecular interaction is expected to be predominant also during film growth, thus, the strong NN-Blatter intermolecular coupling sustains the island growth mode. To evaluate the role played by the interaction with the substrate, an annealing experiment is very useful. The films were subsequentially annealed at 323, 388 and 483 K, respectively, for 30 minutes at each temperature step (Figure 3). The complete desorption at 483 K indicates that the interaction with the substrate is very weak: the molecules are physiosorbed on the $SiO_2$ surface. We also note that the films are remarkably chemical-stable up to 388 K (Figure 3). To investigate the orientation of the molecules on the substrate, we performed NEXAFS spectroscopy that, at the same time, gives information on the unoccupied states and the film structure (Figure 4). We observe a NEXAFS dichroism that indicates that the molecules are not aligned with respect to the substrate.[28] The main peak at around 404 eV is typical of the nitronyl nitroxide NEXAFS spectrum.[29] Its dichroism indicates that the molecules are orientated in a fashion closely resembling that in the single crystal, however, identified by a specific orientation of the film unit cell, with the crystallographic *a*-axis almost perpendicular to the substrate. Thus, the orientation of the molecular plane, not completely aligned with respect to the substrate plane, generates a different image charge for carbon and nitrogen atoms, because of the difference distance from the substrate. This is reflected by the XPS non-rigid shift.

The poor air stability of the NN-Blatter thin films[6] hinders ex-situ investigations traditionally used to directly probe the radical persistence, such as electron paramagnetic resonance spectroscopy. To dispel any doubts on the radical character in the films, besides using the fit procedure, we adopted several strategies: 1) the comparison of the thin film core-level spectra with the same spectra obtained for the powder that show the same features, confirming the stoichiometry argument (see Ref. [6] and the Supporting Information). 2) The comparison of the N 1s core level spectra with N 1s core level spectra obtained as a stoichiometric sum of the single radical thin films, i.e., the NN and the Blatter radical,[8] as shown in Figure 1, that prove the remarkable agreement between the NN-Blatter experimental N 1s core level spectrum and the N 1s core level sum spectrum. 3) The use of ab-initio calculations to further verify that the experimental occupied and unoccupied states correspond to intact diradicals (see below).

This approach clearly shows that it is possible to evaporate the NN-Blatter without degradation. Our goal is understanding why diradicals are extremally difficult to evaporate and show limited air stability, even when they are based on single radicals that are highly robust.

A large variety of novel, stable radicals were synthesized in recent years. We demonstrated that suitable radicals for evaporation are characterized by protection of the spin centers with steric hindrance or very large delocalization of the unpaired electron.[7] This is a fundamental starting point for a successful evaporation. The controlled evaporation in UHV of a radical is more complicated than in the case of close-shell molecules because the space of evaporation parameters, such as pressure, substrate temperature, evaporation temperature, is very limited: on the one hand the evaporation temperature must be high enough to overcome the intermolecular interaction, leading to sublimation, on the other hand, it must be low enough to avoid damaging the radical. Therefore, the temperature window left for a successful evaporation is very narrow. A similar argument also holds for the substrate temperature. In general, the Knudsen cell temperature for radical evaporation in UHV varies in the range between 390 and 420 K (Figure 5). Using higher evaporation temperatures damages the radical. This is a clear difference in comparison with close-shell systems that stand much higher temperatures.[30] In the case of NN-Blatter, the optimized evaporation temperature is significantly lower than in the case of the NN and Blatter radicals taken separately (373 K, versus 393 and 418 K, respectively[8, 10]). Additionally, its film stability is extremely limited in comparison with the one observed for the NN and the Blatter thin films[8, 10] (see Figure S1 in the supporting information). To figure out which molecular property plays a role, we evaporated a second diradical (BTD-NN, Figure 6) adopting the same procedure as for the NN-Blatter thin films. We have chosen BTD-NN because it is a S= 1/2 diradical that forms weakly antiferromagnetically coupled dimers.[18] The distance between the two NN groups in the molecule is 0.7 nm, and its molecular weight is slightly higher than NN-Blatter.[18] The idea behind this choice is to explore the influence of the magnetic coupling as well as the radical interaction on the film growth and stability (see also Table 1). The core level spectra are shown in Figure 6, together with the powder spectra.

Analogously to what we observed for NN-Blatter, we found that the BTD-NN film C 1s spectrum is characterized by a main line at 285.8 eV due to photoelectrons emitted from the atoms in the aromatic ring and the carbon atoms bound to hydrogen atoms. The shoulder at higher binding energy is due to contributions from the electrons emitted from hetero-bound carbon atoms. The two nitrogen atoms belonging to the nitronyl nitroxide radical have an equivalent chemical environment, thus, we expect a single line in the N 1s spectrum. Indeed, this is the case with the main line at 403 eV, although we also observe signal intensity at lower binding energy, indicating that a small amount of nitronyl nitroxide radicals switched to the imino nitroxide radicals.[17, 31] To shed light on the growth mode of the BTD-NN films, we followed the XPS core level signal of the substrate (Si 2p) by looking at its attenuation upon film deposition (Figure 6). The curve is characterised by a very slow decay. This intensity trend is typical of the Volmer-Weber growth mode, i.e., island growth.[32] The atomic force microscopy ex-situ images are consistent with this observation, clearly showing a film morphology dominated by islands.

Following the same argument as for NN-Blatter, we can conclude that the evaporation of the BTD-NN films is successful, because of the agreement of the film stoichiometry with the theoretical values, the agreement between film and powder spectra and the ab-initio calculations (see below). Also in this case the evaporation temperature (373-383 K) is lower than in the case of the single NN.[8] In the case of the BTD-NN films, we also observed an enhanced beam sensitivity with respect to the single NN radical derivatives.

In order to gain further insight, we carried out a first principles study, based on density functional theory (DFT), of the structural, electronic and magnetic properties of both diradicals in the isolated (i.e., single molecule) and aggregated (i.e., bulk-like) phases. We first investigated the stability and the magnetic order of the single molecules, by means of a set of total-energy-and-forces simulation at fixed values total magnetization $M_T$=0,1,2 $\mu_B$, where $M_T = \int (n_{up} - n_{dw}) d^3r$ is the integral of the magnetization in the cell, and $n_{up}$ and $n_{dw}$ are the spin-up and the spin-down components of the electron charge density, described within the local spin density approximation. Thus, $M_T$=2 $\mu_B$, corresponds to two unpaired electrons with parallel spin (↑↑) and total spin S=1. In the case of $M_T$=0 we distinguish between non-magnetic spin unpolarized system (- -) and antiparallel (↑↓) spin arrangement, through the evaluation of the absolute magnetization $M_A = \int |n_{up} - n_{dw}| d^3r$. Non-magnetic systems have $M_T$= $M_A$=0 $\mu_B$, while and two antiparallel spin systems have $M_T$=0 $\mu_B$ and $M_T$=2 $\mu_B$.

Figure 7a shows the total energy variation ΔE of the isolated molecules, as a function of the magnetic state $M_T$, with respect to the non-magnetic state, assumed as the energy zero reference. For both diradicals all the magnetic phases $M_T$ ≠0 are energetically favoured, even though the case of a single unpaired electron $M_T=1\mu_B$ (↑ -) is the less stable one. This confirms the intrinsic diradical character (i.e., two unpaired electrons) of both molecules. Rather relevant differences hold for the spin alignment of the two systems. In the case of NN-Blatter (dark blue line), the parallel (↑↑) spin ordering is favoured by ~50 meV/mol with respect to the antiparallel (↑↓) one, indicating an intra-molecular spin coupling between the two unpaired electrons. On the contrary, for BDT-NN the parallel and antiparallel spin distributions are energetically equivalent (↑↑ is energetically more stable than ↑↓ only by 6 meV/mol). This corresponds to 2 uncorrelated spin radicals (S=1/2) allocated in the same molecule (one per NN unit).

The origin of this behaviour can be explained by the analysis of the electronic structure. Figure 8a shows the spin-polarized density of states (DOS) of the two diradicals in the parallel (↑↑) spin phase. The magnetic character of the NN-Blatter derives from two single occupied molecular orbitals (labelled $S_1$ and $S_2$). $S_1$ and $S_2$ are π-like conjugated states: the former is mostly localized on the NN unit, the latter is centred on the Blatter part; but both wavefunctions have tails that extend over the phenyl ring connecting the two units. The corresponding spin-density plot ($n_{up}-n_{dw}$) in Figure 8b results to be extended on the entire core of the molecule. This is a fingerprint of the intra-molecular coupling between the two radical subsystems and the overall spin alignment of the diradical. On the contrary, in the case of BDT-NN the magnetic character stems from two orbitals (labelled S) that are degenerate in energy but spatially centred in a separate NN unit. The corresponding spin density (Figure 8b) is localized on the O-N-C-N-O bonds, with no overlap in the central core. This corresponds to two unpaired S=1/2 radicals hosted in the same molecule, with no preference for their relative orientation. This justifies also the same total energy for the parallel and antiparallel distribution, shown in Figure 7a.

A similar analysis has been carried out for the bulk phases. In both cases, the initial structure has been taken from experimental X-ray crystalline structure.[6, 18] Each crystal has a triclinic lattice symmetry and includes four diradical units in the original cell. The NN-Blatter crystal is further stabilized by two dichloromethane ($CH_2Cl_2$) molecules. For both systems we optimized the atomic structure for different values of the total magnetization per cell, corresponding to a non-magnetic, FM and AF long-range intermolecular order. The results about the relative energetic stability are summarized in Figure 7b. In the ground state, the Blatter-NN solid phase exhibits a different short-range (intra-molecular) and long-range (inter-molecular) spin alignment distribution, which corresponds to a $M_T=0\mu_B$/cell and $M_A=8.2\mu_B$/cell. Spins remain parallelly oriented (↑↑) within each single molecule, as discussed above, while they arrange in chains with AF order between next-neighbour molecules, (↑↑ - ↓↓), in agreement with the experimental evidence.[6, 18] The fully parallelly oriented spin configuration, i.e., the FM state (↑↑-↑↑), is energetically less favoured by 41 meV/mol than the AF phase.

BDT-NN crystal may be stabilized in both FM and AF phase at the same total energy, being the AF state only 2 meV/mol more stable than the FM one. In both cases, the absolute magnetization is $M_T=8.5$ $\mu_B$/cell, while $M_T=8.0$ $\mu_B$/cell and $M_T=0$ $\mu_B$/cell, for the FM and AF state, respectively. The weak AF behaviour detected in the experiments can be thus interpreted as a sequence of weak interacting S=1/2 NN radicals.

To confirm this analysis, we simulated the N1s X-ray spectra for the two diradicals (Figure 9), in the single molecule configuration. The XPS spectra of BDT-NN (top panel) is dominated by one single peak relative to the $N_{NN}$ atoms that result to be almost indistinguishable. Solid state packing does not change this picture, in agreement with the experimental results (Figure 5). We conclude that each NN unit, both in the molecular and solid-state phase, remain intrinsically independent and weakly coupled to the rest of the molecule or to the environment (i.e., S=1/2 single radicals). On the contrary, the intramolecular interactions affect also the N1s spectra of the NN-Blatter (bottom panel). Four peaks are clearly visible, the one at higher binding energy corresponds to the NN site, the other three ($N_1$, $N_2$, $N_4$) to the Blatter component, in very good agreement with the analysis of experimental data in Figure 1d.

Finally, we investigated the thermodynamic stability of the two compounds, by evaluating the formation energies of the crystal ($E_{for}$), with respect to the single molecule components. Both systems result to be energetically very stable with $E_{for}$=-1.6 eV/mol -0.9 eV/mol, for BDT-NN and NN-Blatter, respectively. We further checked the effect of the dichloromethane units[6]: we prepared and re-optimized the atomic structure of a second crystal that does not included the $CH_2Cl_2$ molecules.[6] Their presence has no effect on the magnetic order of the system but affect their stability. The formation energy of NN-Blatter reduces to -0.4 eV/mol when dichloromethane is not included in the system.

We have gained at this point a full understanding of the systems that allows us to answer the initial question on why the diradicals are difficult to evaporate. Our results show that the strength of the antiferromagnetic interaction does not play a role, since we observe a very similar behaviour for both radicals that have different antiferromagnetic interaction strength (strong versus weak). The distance between the radical sites and the spin do not influence (in this case 1.0 nm versus 0.7 nm, and 1 versus ½, for the NN-Blatter and BTD-NN, respectively). The thermodynamics results are the key. The two diradicals assemble in stable aggregates (high $E_{for}$), which includes both the initial powder (before evaporation) and the final films (after deposition). On the one hand, to evaporate the molecules it is necessary to furnish enough thermal energy (i.e., increasing the temperature) to release each molecule to its gas phase from the initial crystalline powder. On the other hand, temperature cannot be arbitrary increased to avoid compromising the structural integrity of the molecules. This restricts the evaporation to a very tiny temperature window, as discussed for the experiments. Once the molecules reach the target substrates, they easily tend to re-aggregate growing in compact island rather than in ordered planar films (Figures 2 and 6). Kinetics most probably affects the morphology and the size rather than the formation of the molecular islands.

Instead, the increased film instability in air depends on the number of radical sites (radical versus diradical) when comparing the same radical in single or diradical configuration. From the comparison between NN-Blatter and BTD-NN thin films, we also note that the delocalisation of the unpaired electron play an important role. The NN-Blatter thin films are more robust than the BTD-NN thin films. If we look at the molecular structure, BTD-NN carries two NN, with the two unpaired electrons delocalised over the two NO groups, one in each radical. NN-Blatter on the contrary carries only one NN radical and one Blatter radical. The delocalisation of the unpaired electron in the Blatter radical is quite extended. Already taking the single radicals into account, the films of the Blatter radical derivatives are more stable than those of the NN derivatives.[10, 17] This difference is strongly enhanced when two NN radicals are present in the same molecules.

**Conclusions**

We have demonstrated that it is possible to evaporate diradical in a controlled environment obtaining thin films in which the diradical character is preserved. However, evaporation represents a challenge, also in the case of the presence of very stable single radicals. We have explored the parameters that play a role in this phenomenon. Bulk formation thermodynamics and delocalisation of the unpaired electrons play the major role. The higher the formation energies of the crystal, the more difficult is the evaporation of intact radicals. The larger the delocalization, the more stable is the film exposed to air. The evaporation of different radicals, also with a larger number of radical sites can be successfully addressed considering our findings.

**Methods**

**Experimental Section**

Thin films were prepared in situ under UHV conditions by OMBD using a Knudsen cell. Our measurements showed that the residual powder in the cell that underwent four evaporation cycles led to some minor degradation in the films, therefore, we grew the films using the same residual powder in the Knudsen cell for not more than three successive evaporation cycles. Native $SiO_2$ grown on single-side polished n-Si(111) wafer was used as the substrate for all thin films. The wafers were cleaned in an ultrasonic bath by immersion in ethanol and acetone for one hour each and annealed at around 500 K for several hours. Cleanness was verified by XPS. Nominal films thicknesses were calculated from the attenuation of the substrate signal. The NN-Blatter deposition rate was 0.03 nm/min. The substrate was kept at room temperature during deposition. The thin films were grown by direct deposition, unless specified differently in the text. The UHV system was composed of a dedicated OMDB chamber connected to an analysis chamber ($2\times10^{-10}$ mbar base pressure) in which the XPS measurements were conducted. It is equipped with a monochromatic Al K$\alpha$ source (SPECS Focus 500) and a SPECS Phoibos 150 hemispherical electron analyser. Survey spectra were measured at 50 eV pass energy and individual core level spectra at 20 eV pass energy. Both were subsequently calibrated to the Si 2p signal at 99.8 eV. To minimise potential radiation damage, only freshly prepared films were measured, and radiation exposure was minimised. For measurements probing stability, beam exposure was further limited after air exposure in order to attribute the damage exclusively to the degradation by air exposure. This results in a worse signal to noise ratio in those spectra.

NEXAFS measurements were performed at the third-generation synchrotron radiation source BESSY II at the Low-Dose PES end station installed at the PM4 beamline ($E/\Delta E = 6000$ at 400 eV). This end station was equipped with a similar setup as the one described above. The same calibrated Knudsen cells used to grow the films for the XPS measurements were mounted to a preparation chamber attached to the NEXAFS measuring chamber to reproduce the same preparation conditions, as for XPS. The measurements were carried out in multibunch hybrid mode (ring current in top up mode = 300 mA). The spectra were measured in total electron yield and normalised with the clean substrate signal as well as the ring current. Subsequently they were scaled to give an equal absorption edge jump.

Atomic force microscopy was measured under ambient conditions with a Digital Instruments Nanoscope III Multimode microscope using tapping mode.

**Computational Section**

Single molecules and crystalline systems have been simulated by using first principles approaches based on DFT, as implemented in the Quantum-Espresso suite of codes.[33] Spin-unrestricted geometry optimizations were performed at the PBE-GGA level,[34] and van der Waals corrections were included within the semiempirical method proposed by S. Grimme (DFT-D2).[35] The electronic structure is described by using B3LYP hybrid-functional in order to correct the DFT deficiencies in reproducing the bandgap, without further atomic relaxation on the optimized PBE structures. Spin degrees of freedom are treated within the local spin density approximation. Atomic potentials are described by ultrasoft pseudopotential as available in the SSSP library.[36] Single particle wavefunctions (charge) are expanded in planewaves up to a kinetic energy cutoff of 30 Ry (300 Ry), respectively. The first Brillouin zone of bulk systems is sampled with a (2x2x2) k-point grid; center-zone Γ-point is used in the case of single molecules. All structures were fully relaxed until forces on all atoms become lower than 0.03 eV/Å.

The N 1s core level spectra were calculated in the pseudopotential framework using the final state theory.[37] This approach provides only the relative shift between the core level binding energies of inequivalent atoms, while their absolute value is not defined. The choice of the reference does not change the analysis.

**Acknowledgements**

The authors would like to thank Helmholtz-Zentrum Berlin (HZB) for providing beamtime at BESSY II (Berlin, Germany), Hilmar Adler, and Elke Nadler for technical support. Financial support from HZB, and German Research Foundation (DFG) under the contract CA852/11-1 is gratefully acknowledged. We thank the National Science Foundation (NSF), Chemistry Division for support of this research under Grants No. CHE-1665256 (A.R.) and CHE-1955349 (A.R.).

**Notes**

The authors declare no competing financial interest.

†These authors contributed equally.

Figure 1

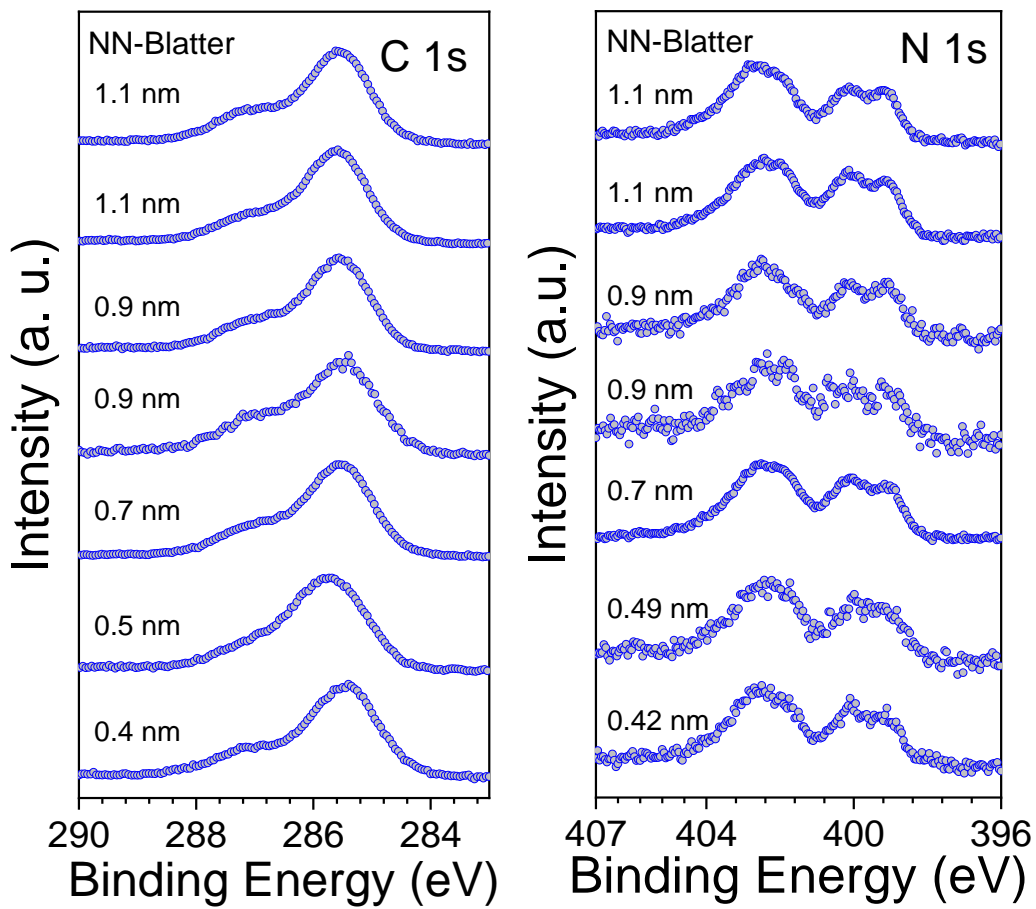

a)

b)

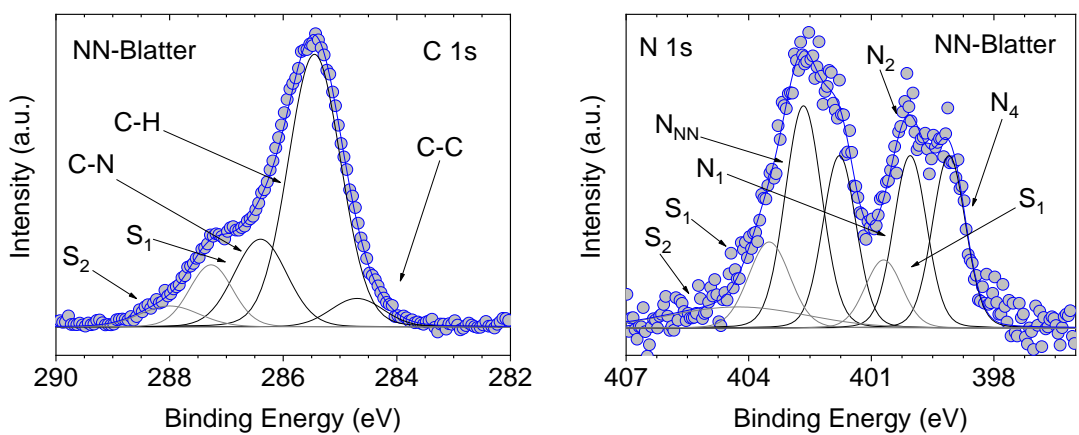

c)

d)

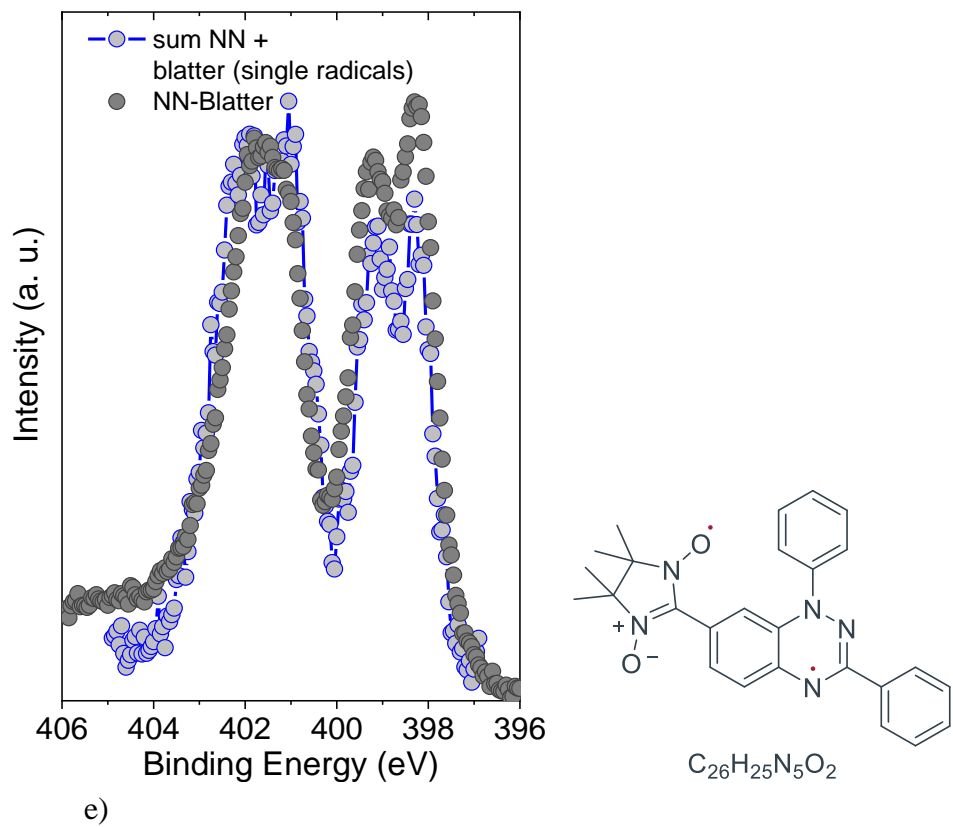

Figure 1 Top panel: NN-Blatter a) C 1s and b) N 1s thickness-dependent core-level spectra of the NN-Blatter thin films, thickness as indicated. NN-Blatter c) C 1s and d) N 1s spectra with their best fit for the 0.4 nm thick film. e) N 1s core level spectrum obtained as a sum of the core level spectra of the single radicals (NN + Blatter radical[8, 10]) compared to the NN-Blatter N1s core level spectrum as in b). The NN-Blatter structure is also shown.

Figure 2

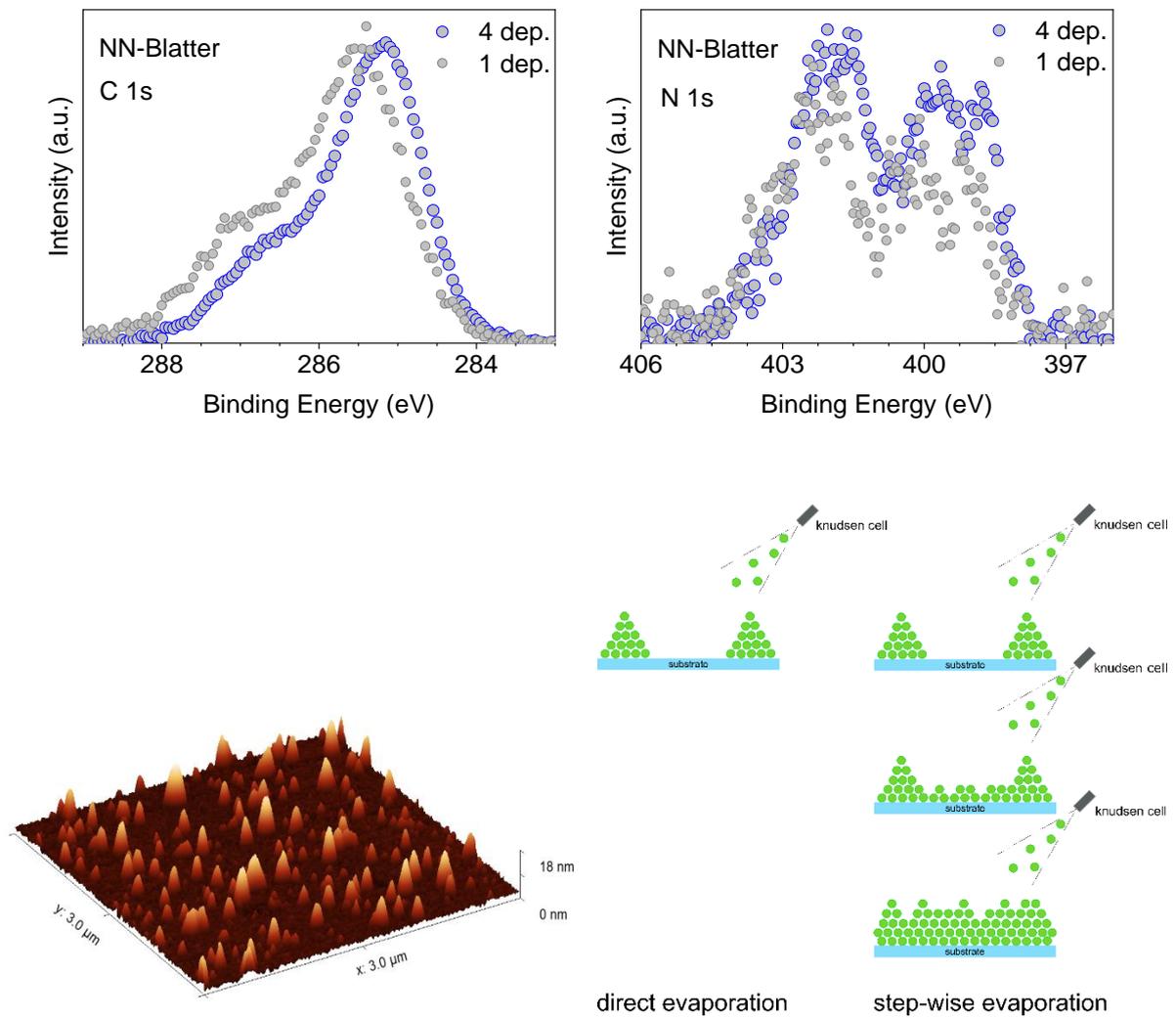

Figure 2. NN-Blatter a) C 1s and b) N 1s core level spectra: direct versus stepwise deposition (substrate kept at room temperature in both cases). c) Sketch of the growth mode in the two cases. d) 3 μm x 3 μm atomic force image image of a NN-Blatter thin film (direct evaporation).

Figure 3

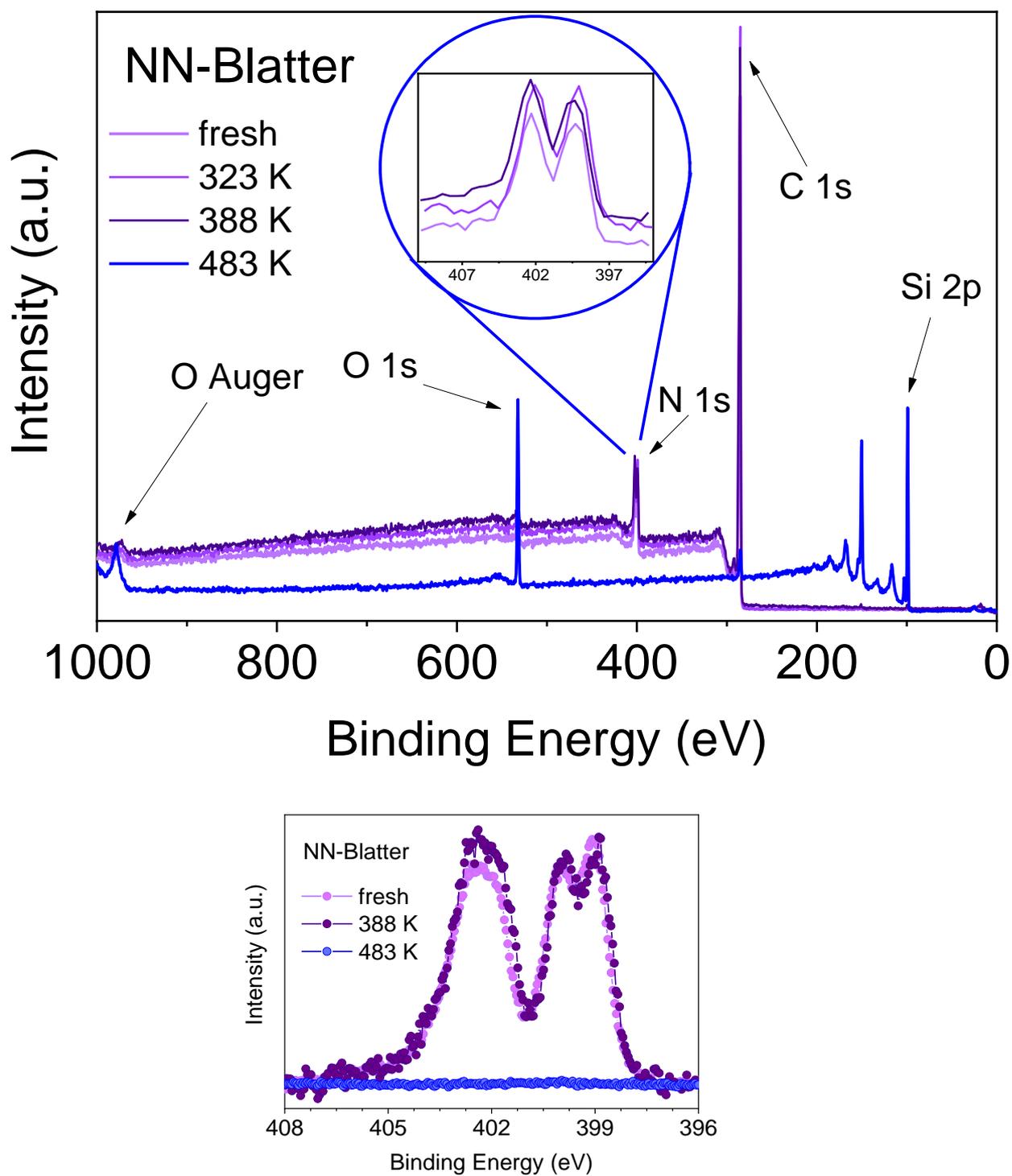

Figure 3. NN-Blatter XPS survey spectra together with the detailed N 1s core level spectra at different annealing temperatures, as indicated.

Figure 4

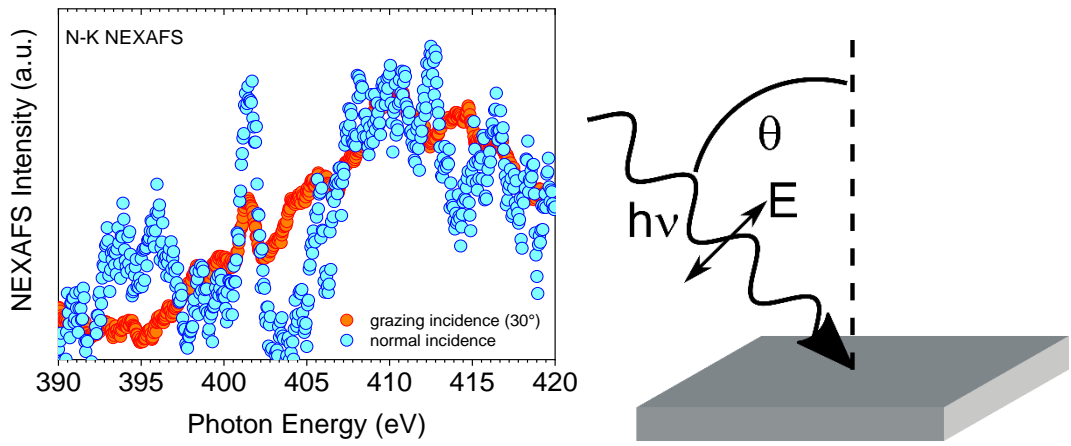

Figure 4. NN- Blatter N-K NEXAFS spectra with the geometry of the experiment.

Figure 5

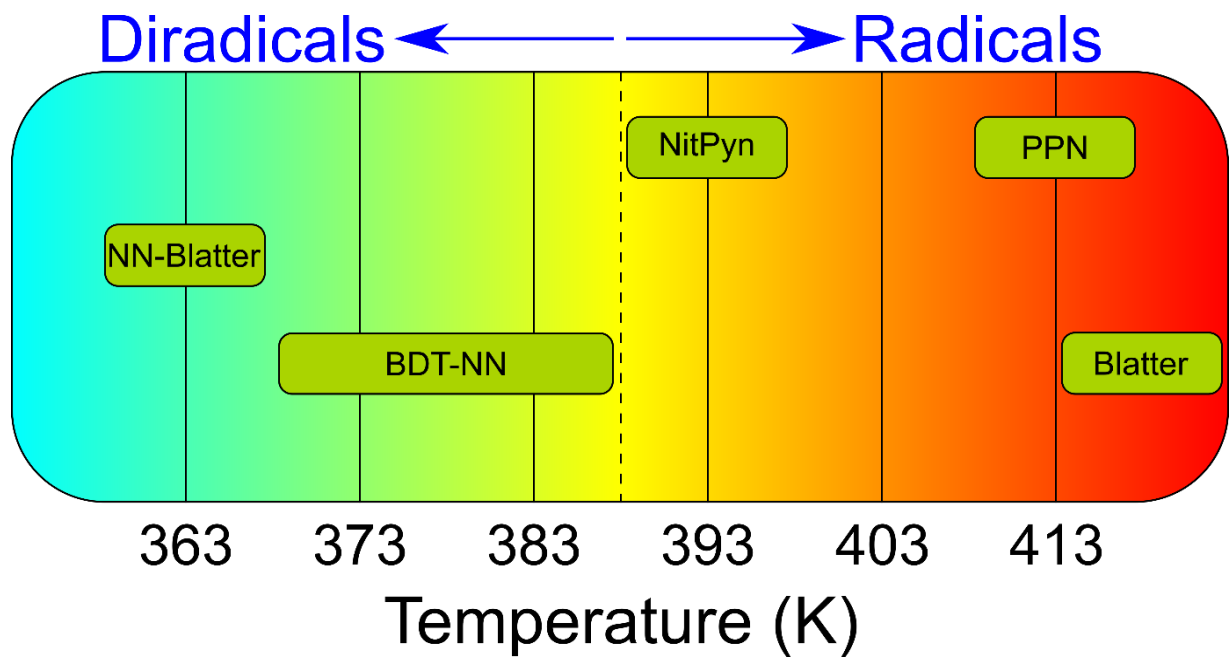

Figure 5. Schematic diagram of the evaporation temperature range for several diradicals and radicals.

Figure 6

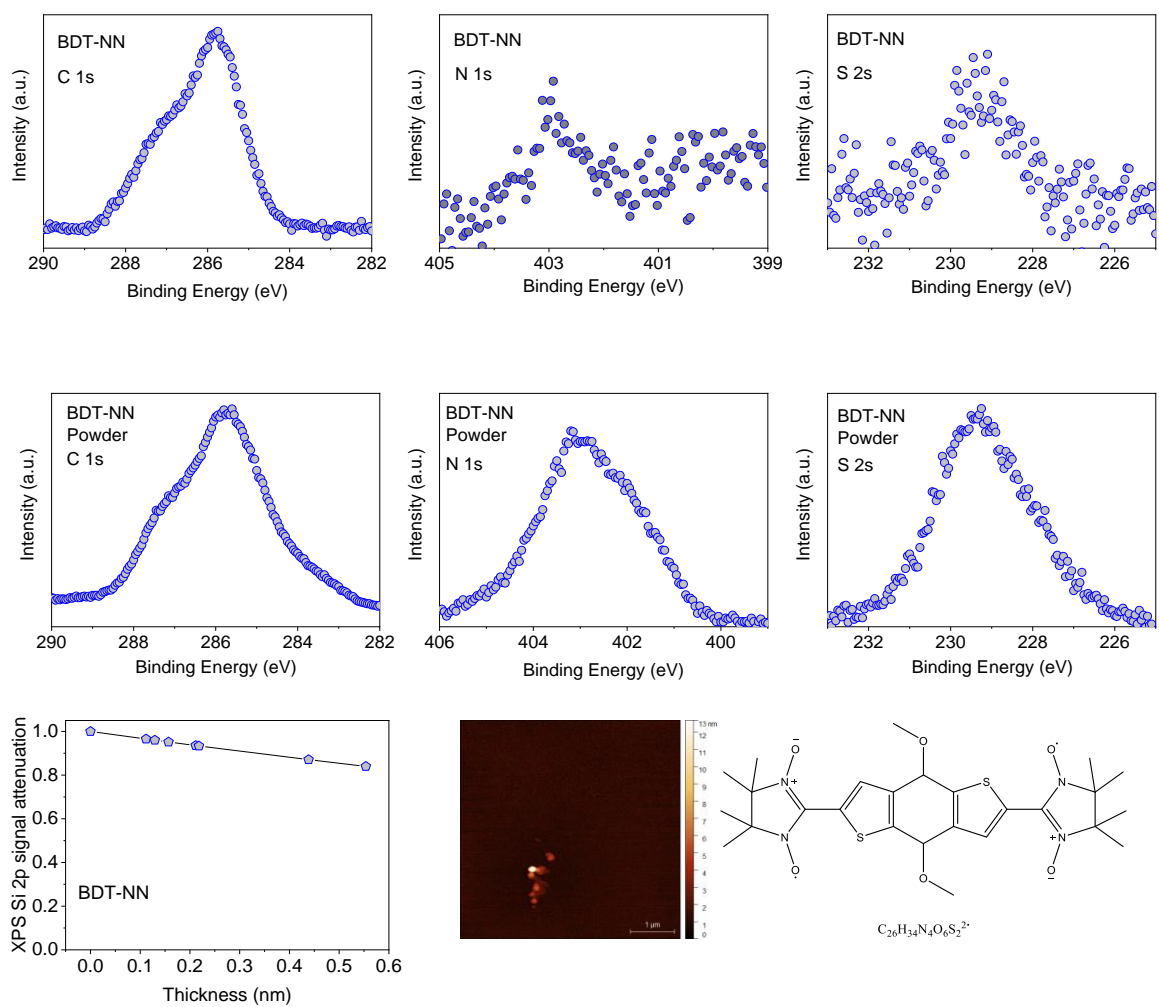

Figure 6. BTD-NN a) C 1s, b) N 1s and c) S 2s core-level spectra of the BTD-NN thin films, compared to the correspondent spectra of the powder, d)-f), as indicated. Attenuation of the Si 2p XPS signal, normalized to the corresponding saturation signal, as a function of film nominal thickness, deposition at RT. The BTD-NN structure is also shown.

Figure 7

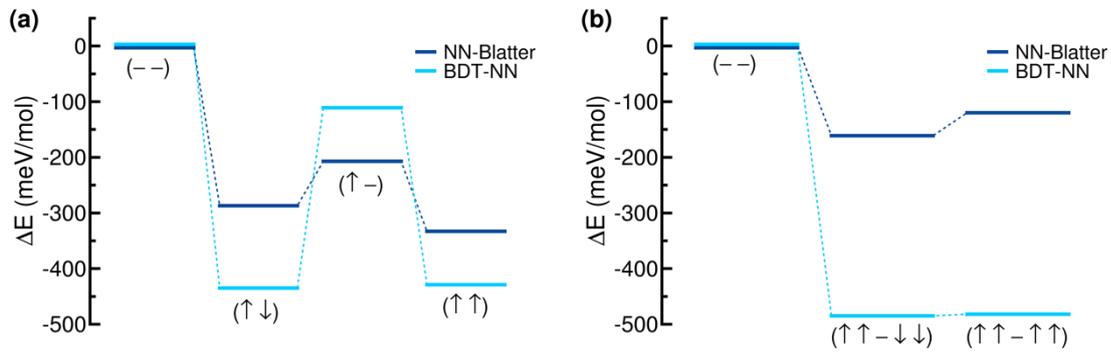

Figure 7. Total energy difference ΔE of (a) single molecules and (b) bulk crystals, as a function of the total magnetization, with respect to the non-magnetic (- -) phase. In panel (a) labels (↑↓), (↑ -), (↑↑) describe the single molecule (intra-molecular) spin alignment and correspond to $M_T$=0, 1, and 2$\mu_B$/mol, respectively; in panel (b) (↑↑-↓↓) and (↑↑-↑↑) correspond to AF and FM long-range (inter-molecular) spin alignment, and correspond to $M_T$=0, and 8$\mu_B$/cell, respectively.

Figure 8

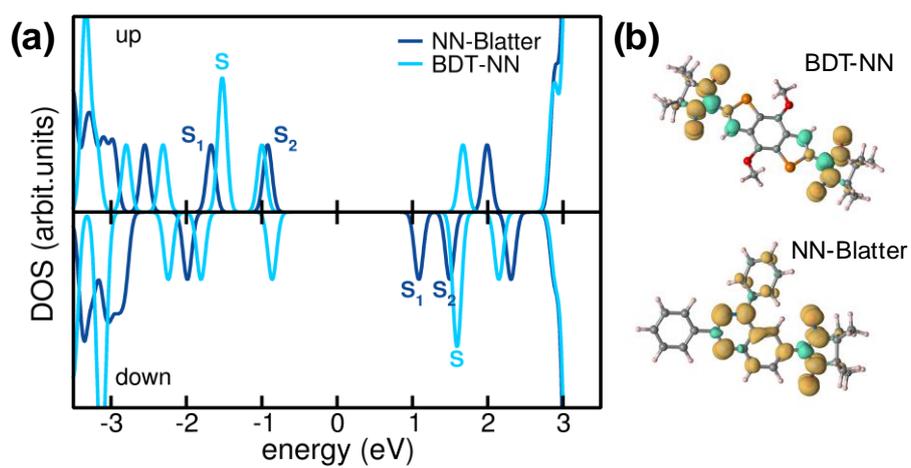

Figure 8. (a) Spin-polarized DOS plot for NN-Blatter (dark blue) and BDT-NN (light cyan) diradicals (single molecule). The corresponding spin-density isosurface plots are shown in panel (b).

Figure 9

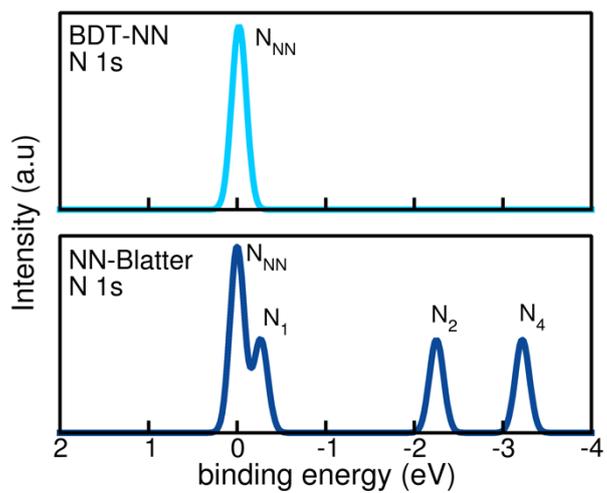

Figure 9. Simulated X-Ray N1s spectra of BDT-NN (upper panel) and NN-Blatter (bottom panel) in the isolated molecule configuration. NN peak is assumed as zero energy reference for all spectra.

Table 1

|  | BDT-NN | NN-Blatter |
|---|---|---|
| Formula | $C_{26}H_{32}N_4O_6S_2$ | $C_{26}H_{25}N_5O_2$ |
| Formula weight (g/mol) | 560.67 | 481.97 |
| Density (g/cm$^3$) | 1.332 | 1.363 |
| Spin | 1/2 | |
| $J/k_B$ (K) | -26 | -14 |
| Magnetic interaction | weak AFM | Strong AFM |
| Intramolecular ordering | dimer | 1D chains |
| Evaporation Temperature (K) | 373 – 383 | 363 |